\documentclass[twocolumn,10pt,pre, aps,superscriptaddress,showpacs]{revtex4-1}

\usepackage[utf8]{inputenc}
\usepackage[T1]{fontenc}
\usepackage{subfigure}

\usepackage{dcolumn}
\usepackage{amsmath,amssymb}
\usepackage{bm}
\usepackage{color}
\usepackage{latexsym}
\usepackage{psfrag,graphicx}
\usepackage{color}
\usepackage[english]{babel}
\usepackage{latexsym}
\usepackage{epsf} 
\usepackage{amsfonts}
\usepackage{natbib}
\usepackage{epstopdf}
\usepackage{appendix}
\usepackage{changes}
\usepackage{siunitx}

\definecolor{mygrey}{gray}{0.35}
\definecolor{myblue}{rgb}{0.2,0.2,0.8}
\definecolor{myzard}{cmyk}{0,0,0.05,0}
\definecolor{mywhite}{rgb}{1,1,1}
\definecolor{myred}{rgb}{1,0.,0.3}
\definecolor{MATblue}{rgb}{0 0.4470 0.7410}
\definecolor{MATorange}{rgb}{0.8477 0.3242 0.0977}
\definecolor{MATgreen}{rgb}{0 .75 0}

\usepackage[colorlinks=true,citecolor=myblue,linkcolor=myred]{hyperref}

\def\be{\begin{equation}}
\def\ee{\end{equation}}
\def\ba{\begin{align}}
\def\enda{\end{align}}
\def\bi{\begin{itemize}}
\def\ei{\end{itemize}}

\def\beq{\begin{equation}}
\def\beq{\begin{equation}}
\def\eeq{\end{equation}}
\def \bea{\begin{eqnarray}}
\def \eea{\end{eqnarray}}

 \def\ee{\mathord{\rm e}}

\begin{document}

\title{Enhanced laser-driven proton acceleration with gas-foil targets}

\author{Dan Levy}
\affiliation{Department of Physics of Complex Systems, Weizmann Institute of Science, Rehovot 76100, Israel} \affiliation{Laboratoire d'Optique Appliqu\'ee, \'Ecole Polytechnique-ENSTA- CNRS, Institut Polytechnique de Paris, 91761 Palaiseau Cedex, France}

\author{Xavier Davoine}
\author{Arnaud Debayle}
\author{Laurent Gremillet}
\affiliation{Universit\'e Paris-Saclay, CEA, LMCE, 91680 Bruy\`eres-le-Ch\^atel, France}

\author{Victor Malka}
\affiliation{Department of Physics of Complex Systems, Weizmann Institute of Science, Rehovot 76100, Israel}
\affiliation{Laboratoire d'Optique Appliqu\'ee, \'Ecole Polytechnique-ENSTA- CNRS, Institut Polytechnique de Paris, 91761 Palaiseau Cedex, France}

\begin{abstract}
We study numerically the mechanisms of proton acceleration in gas-foil targets driven by an ultraintense femtosecond laser pulse. The target consists of a near-critical-density hydrogen gas layer of a few tens of microns attached to a solid carbon foil with a contaminant thin proton layer at its back side. 
Two-dimensional particle-in-cell simulations show that, at optimal gas density, the maximum energy of the contaminant protons is increased by a factor of $\sim 4$ compared to a single foil target. This improvement originates from the near-complete laser absorption into relativistic electrons in the gas. Several energetic electron populations are identified, and their respective effect on the proton acceleration is quantified by computing the electrostatic fields that they generate at the protons' positions.
While each of those electron groups is found to contribute substantially to the overall accelerating field, the dominant one is the relativistic thermal bulk that results from the nonlinear wakefield excited in the gas, as analyzed recently by Debayle \emph{et al.} [New J. Phys. \textbf{19}, 123013 (2017)]. Our analysis also reveals the important role of the neighboring ions in the acceleration of the fastest protons, and the onset of multidimensional effects caused by the time-increasing curvature of the proton layer.
\end{abstract}

\maketitle

\section{Introduction}
\label{sec:intro}

As first demonstrated two decades ago~\cite{snavely_intense_2000, clark_measurements_2000, maksimchuk_forward_2000}, the interaction of an ultraintense ($I_L\gtrsim 10^{18}\,\rm Wcm^{-2}$) laser pulse with a thin ($\sim 1\,\rm \mu m$) foil target results in the formation of an energetic ion beam through the so-called target normal sheath acceleration (TNSA) mechanism \cite{wilks_energetic_2001}.
Stimulated by a great variety of applications, ranging from radiography \cite{borghesi_electric_2002} and warm dense matter \cite{patel_isochoric_2003} to nuclear physics \cite{ledingham_applications_2003} and proton therapy~\cite{malka_principles_2008}, vast research effort has since been devoted to improving the maximum energy, flux and collimation of the ion beam. In the meantime, other acceleration mechanisms have been proposed, such as radiation pressure acceleration~\cite{macchi_laser_2005,schlegel_relativistic_2009}, light sail acceleration~\cite{esirkepov_highly_2004, robinson_radiation_2008, klimo_monoenergetic_2008, macchi_light_2009,qiao_dominance_2012}, relativistic self-induced transparency acceleration \cite{dhumieres_proton_2005, esirkepov_laser_2006}, breakout afterburner~\cite{yin_monoenergetic_2007, yin_three-dimensional_2011}, collisionless shock acceleration \cite{denavit_absorption_1992, silva_proton_2004, fiuza_weibel-instability-mediated_2012,haberberger_collisionless_2012}, or magnetic vortex acceleration~\cite{fukuda_energy_2009, nakamura_high-energy_2010}. However promising these mechanisms may appear to be, they generally require stringent laser and target properties, rendering their experimental realization very challenging. Therefore, due to its simplicity and robustness, TNSA remains by far the most widely used method for laser-based ion acceleration, motivating extensive optimization efforts concerning target geometry and composition.

TNSA proceeds from the strong electrostatic fields set up at the target surfaces by the laser-driven hot electrons leaking out into vacuum, and forming negatively charged sheaths. Those fields act to reflect the hot electrons back into the target, while driving the outward expansion of the surface ions. Their strength scales as $E_x \simeq k_B T_h/e\, \mathrm{max}(\lambda_{Dh},L_n)$, with $e$ being the elementary charge, $k_B$ the Boltzmann constant, $T_h$ the ``temperature'' of the hot electrons, $\lambda_{Dh}$ their local Debye length, and $L_n$ the local ion density scale length~\cite{mora_plasma_2003}. The sheath field is generally stronger at the target back side because it is not directly perturbed by the laser, and so is characterized by a sharper density gradient ($L_n \ll \lambda_{Dh}$). In laser-solid interactions, the energy of the hot electrons is usually comparable with the ponderomotive potential, $T_h \simeq m_e c^2 \sqrt{1+a_0^2/2}$, where $a_0 = eE_0/m_e c \omega_0$ is the dimensionless laser field ($\omega_0$ is the laser frequency, $m_e$ the electron mass, $c$ the velocity of light) \cite{wilks_absorption_1992}. For typical values $T_h\sim 1\,\rm MeV$, $\lambda_{Dh} \sim 1\, \rm \mu m$, electrostatic fields of a few $10^6\,\rm V \mu m^{-1}$ are generated, readily ionizing the hydrogen-rich surface contaminants. Due to their largest charge-to-mass ratio, the surface protons react the fastest to the sheath field and are accelerated to the highest velocities. Apart from some aspects specific to hot-electron generation, the interplay of hot and bulk electrons and the usually multi-layered and/or multi-species composition of the targets, TNSA is equivalent to the standard problem of plasma expansion into vacuum \cite{gurevich_self-similar_1966, crow_expansion_1975, mora_plasma_2003}.

Since the accelerating field ($E_x \simeq k_B T_h/\lambda_{Dh} \propto \sqrt{n_h T_h}$) increases with the temperature ($T_h$) and density ($n_h$) of the hot electrons, various target configurations have been designed to maximize these parameters for a given laser drive. Among the explored setups, double-layer targets made of a thin solid foil preceded by a plasma layer of near-critical electron density offer particularly encouraging prospects, with up to threefold enhancement in cutoff ion energies being reported \cite{nakamura_interaction_2010,margarone_laser-driven_2012,sgattoni_laser_2012,wang_efficient_2013,passoni_energetic_2014,bin_ion_2015,passoni_toward_2016,prencipe_development_2016,bin_enhanced_2018,ma_laser_2019}. Experimentally, such near-critical plasma layers have been realized using nanostructured carbon foams \cite{passoni_energetic_2014,prencipe_development_2016,bin_ion_2015,bin_enhanced_2018,ma_laser_2019}, nanospheres \cite{margarone_laser-driven_2012} and even bacteria \cite{dalui_bacterial_2014}. 

In this paper, using particle-in-cell (PIC) simulations, we investigate numerically the ion acceleration processes arising in double-layer (or ``gas-foil'') targets comprising a near-critical plasma layer much thicker than the solid-density foil. Specifically, we consider the case of a $2\,\rm \mu m$ thick copper foil target in contact with a $25\,\rm \mu m$ hydrogen gas layer such that when fully ionized it turns into a plasma of the order of the critical density. Our work extends previous simulation studies on double-layer targets, which have so far dealt with shorter and denser foam coatings as first layers \cite{nakamura_interaction_2010,sgattoni_laser_2012}, or with ultrathin foils as second layers, thus giving rise to possibly different ion dynamics \cite{wang_efficient_2013, bin_enhanced_2018, ma_laser_2019}. The originality of our approach is to discriminate between the contributions from the various charged layers of the system to the ion-accelerating electric field. Notably, we are able to evaluate the contributions of the hot electrons as a function of their position and energy.

The paper is organized as follows. After detailing the physical and numerical parameters of the PIC simulations, we describe the sequential processes responsible for TNSA in gas-foil targets. A simulation scan over the gas density highlights the critical role of the gas in determining the laser energy absorption into hot electrons and the final proton energy. We then identify the different sources of the ion acceleration. This is done first by resolving the total electrostatic field seen by the outermost protons into the fields created by different plasma regions, and then by pinpointing the electron groups that account for most of the accelerating field. Finally, we discuss the impact of the finite transverse size of the sheath field and of the ensuing curvature of the ion front on the process.

\section{Simulation parameters}
\label{sec:numerical_setup}

Our PIC simulations are performed in 2D3V geometry (2D in physical space, 3D in momentum space) using the fully electromagnetic and relativistic \textsc{calder} code \cite{lefebvre_electron_2003}. The laser pulse is modeled with a Gaussian envelope of $30\,\rm fs$ FWHM duration and $3\,\rm \mu m$ FWHM transverse size. It has a $0.8\,\rm \mu m$ wavelength, a  $3\times 10^{20}\,\rm W\,cm^{-2}$ peak intensity (corresponding to a dimensionless field strength $a_0=12$), propagates along the $+x$ direction and is polarized along the (in-plane) $y$ axis.

The reference target, irradiated at normal incidence, is a 2-$\rm \mu m$-thick solid carbon foil, coated on its rear side with a $6.3\,\rm nm$-thick hydrogen layer that models the experimental surface contaminants. The initially neutral carbon and hydrogen atoms are initialized with a particle density of $57.6\,n_c$, where $n_c = 1.74\times 10^{21}\,\rm cm^{-3}$ is the (non-relativistic) critical density associated with the laser wavelength. 
In gas-foil targets, a hydrogen gas layer is added on the front side of the foil, consisting of a $24.5\,\rm \mu m$-long plateau of electron density varying in the $0.1-2\,n_c$ range. In all simulations, the laser pulse is focused at the front side of the solid foil. Through impact and field ionization, the carbon ions in the foil target reach an average charge state of +4 at the end of the simulation, corresponding to a $\sim 230 n_c$ average electron density.

The foil is placed at the center of the $L_x\times L_y = 63.7 \times 63.7\,\rm \mu m^2$ simulation domain. The cell sizes are $\Delta x=\Delta y=0.05 c/\omega_0$ ($6.4\,\rm nm$) and the time step is $\Delta t=0.035 \omega_0^{-1}$ ($0.015\,\rm fs$). The carbon ions are represented by 180 macro-particles per cell, leading to an average number of 720 macro-electrons per cell after ionization. The hydrogen gas and surface layers are modeled by 16 and 450 macro-particles per cell, respectively. Current projection and field interpolation are performed using Esirkepov's \cite{esirkepov_exact_2001} and Sokolov's \cite{sokolov_alternating-order_2013} schemes, together with a 4th order weight factor to quench numerical heating.

The start time of proton acceleration in the foil target depends on the dynamics of the laser-accelerated electrons in the gas layer, and so may vary between simulations. Consequently, the time origin ($t=0$) is defined henceforth as the moment when the fastest protons from the foil have attained an energy of $0.01\,\rm MeV$.

\begin{figure}[t]
\centering
\includegraphics[width=0.8\columnwidth]{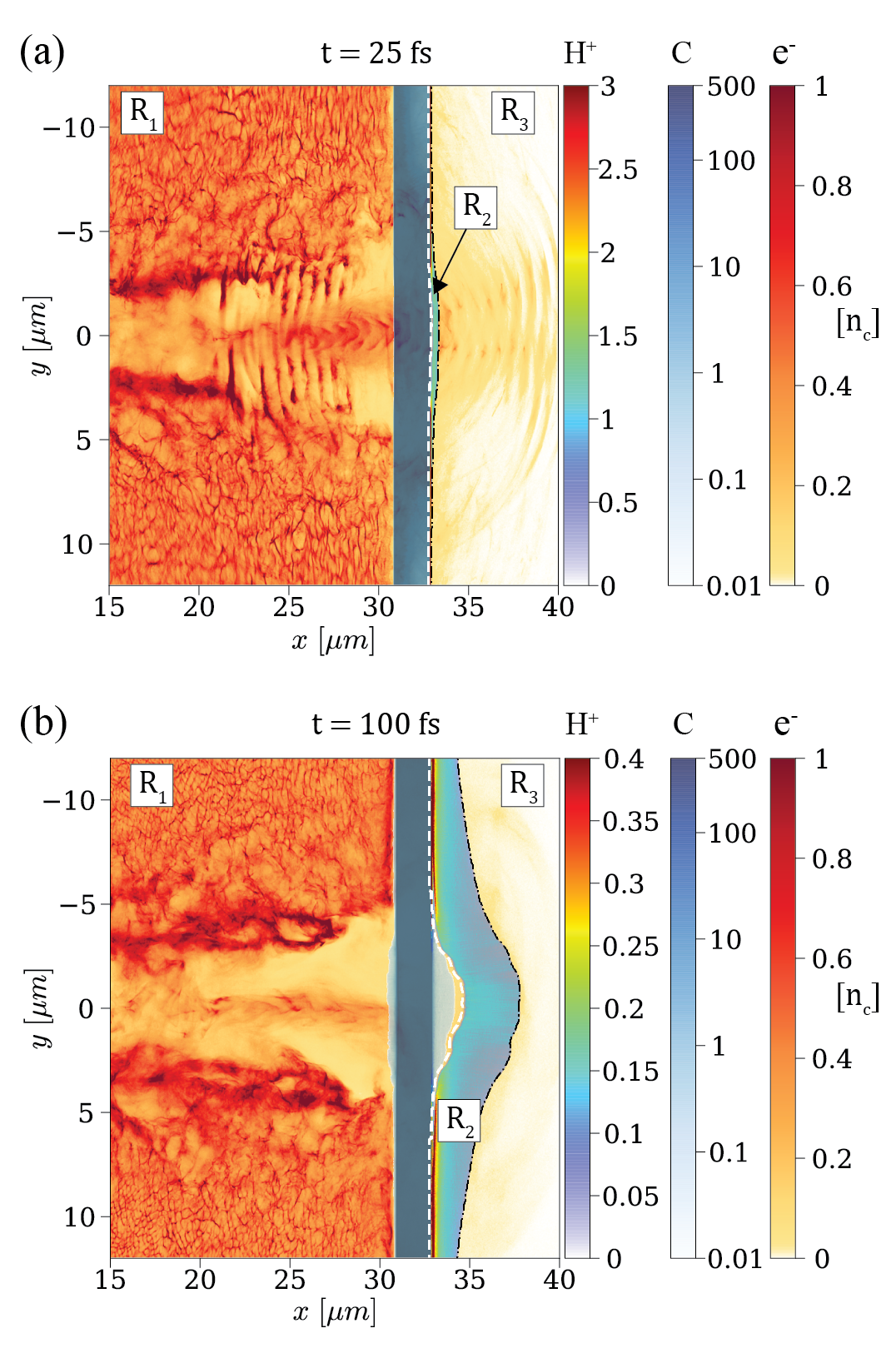} 
\caption{Ion acceleration in the gas-foil target: 2D density maps (in $n_c$ units) of the protons, carbon ions and electrons from the gas layer at two successive times. The dashed lines delineate the boundaries between the three regions $R_1,R_2,R_3$ as discussed in Sec.~\ref{subsec:contrib_S}. (a) At $t=25\,\rm fs$, the $3\times 10^{20}\,\rm W\,cm^{-2}$, $30\,\rm fs$ laser pulse coming from the left has traversed the $0.5\,n_c$ gas layer, where it has generated a strong current of fast electrons, and hit the solid carbon foil. The thin proton layer starts to be accelerated by the sheath field induced by the fast electrons escaping into vacuum, yet it is still attached to the back side of the foil.
(b) At $t=100\,\rm fs$, the proton layer has expanded a few microns; its central part has detached from the foil, and its outer front surface has developed a marked concave shape.}
\label{fig:density_maps}
\end{figure}

\section{Enhanced ion acceleration in gas-foil targets}
\label{sec:enhanced_acceleration}

\subsection{Phenomenology} \label{subsec:phenomenology}

The ion acceleration process taking place in a gas-foil target is illustrated by the two snapshots shown in Fig.~\ref{fig:density_maps}. The initial density of the gas layer is $n_e=0.5\,n_c$, giving rise to maximum proton acceleration from the foil (see below). In each panel are overlaid the density maps of the foil's carbon ions, surface contaminant protons and electrons. Only those electrons originating from the gas are displayed since, as later discussed, those from the foil layer contribute only weakly to the proton final energy. The top panel [Fig.~\ref{fig:density_maps}(a)] shows the early stage of ion acceleration ($t=25\,\rm fs$), when the laser pulse has propagated through the undercritical plasma and started interacting with the solid foil. A cloud of fast electrons (light yellow), modulated by the laser ponderomotive force, is breaking out through the rear side of the carbon foil (dark blue). There ensues a (mainly) longitudinal electrostatic field that sets into motion the contaminant protons (cyan). As a result of the laser interaction, a partially electron-depleted channel has formed in the hydrogen gas, with a filament along its axis made of protons and hot ($> 2\,\rm MeV$) electrons. These electrons are energized by various mechanisms during the laser propagation through the gas, as described in the next section. The high electron energy gain (compared to laser-solid interaction) and the large volume of interaction entail almost complete laser absorption and, therefore, a substantial increase in the electrostatic field set up at the target rear side.

The bottom panel [Fig.~\ref{fig:density_maps}(b)] depicts a later stage of the ion acceleration ($t=100\,\rm fs$). The transmitted fraction of the laser pulse has then already reflected off the solid foil, and the proton layer, now completely separated on axis from the carbon ions, exhibits a substantial curvature. As the proton layer expands, the outermost (higher-energy) front protons experience a diminishing electric field until reaching saturation in energy.

\subsection{Effect of the gas on laser absorption and proton energies}

\begin{figure}[t]
\centering
\includegraphics[width=0.8 \columnwidth]{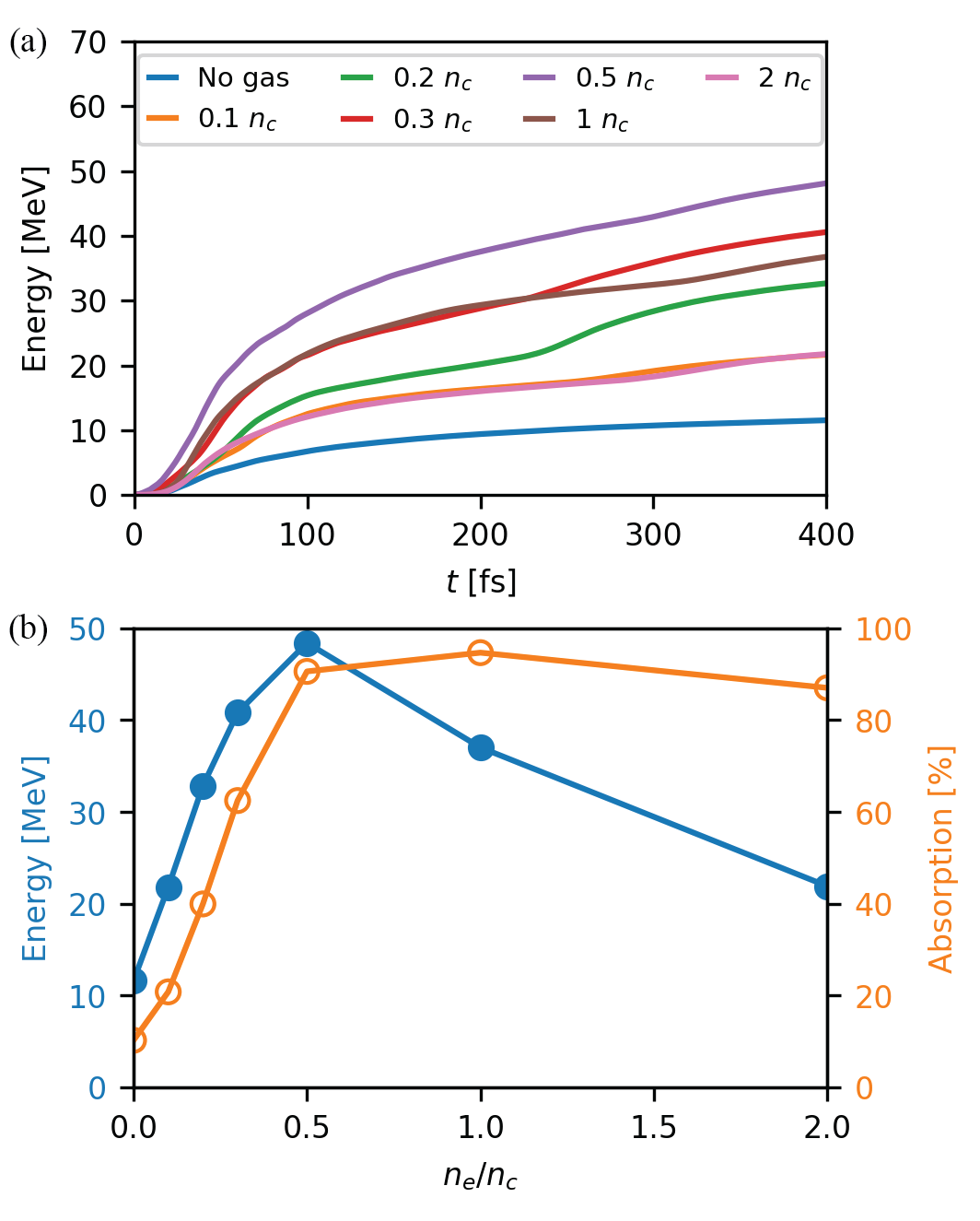}
\caption{(a) Temporal evolution of the maximum proton energy for different gas densities.
(b) Maximum proton energy at the end ($t=400\,\rm fs$) of the simulation (blue) and absorbed laser fraction (orange) as a function of gas density.}
\label{fig:E_max_absorption}
\end{figure}

The efficiency of proton acceleration from gas-foil targets closely depends on the electron density of the gas layer. This is illustrated in Fig.~\ref{fig:E_max_absorption}(a), which plots the temporal evolution of the maximum proton energy for different gas densities. All curves show a rapid initial rise followed by a slower growth phase. The initial growth rate of the energy tends to increase with the electron gas density up to $n_e \simeq 0.5\,n_c$, and to decrease with $n_e$ beyond this value.
The simulations are stopped at $t=400\,\rm fs$, that is, when the protons reach the right-hand side of the domain (after traveling a distance of $\sim 31\,\rm \mu m$) in the optimal $n_e=0.5\,n_c$ case. None of the curves has then reached complete saturation; this is a common feature of the reduced 2D geometry, which causes overestimation of the hot electron density, and hence of the efficiency and timescale of the proton acceleration \cite{liu_three_2013}. A more accurate description would require 3D simulations, yet currently outside the reach of our computational resources. Despite this limitation, we expect the ordering of the different curves plotted in Fig.~\ref{fig:E_max_absorption} to be representative of the scaling with density that might be obtained in a real-world 3D scenario.

\begin{figure*}[t]
\centering
\includegraphics[width=\textwidth]{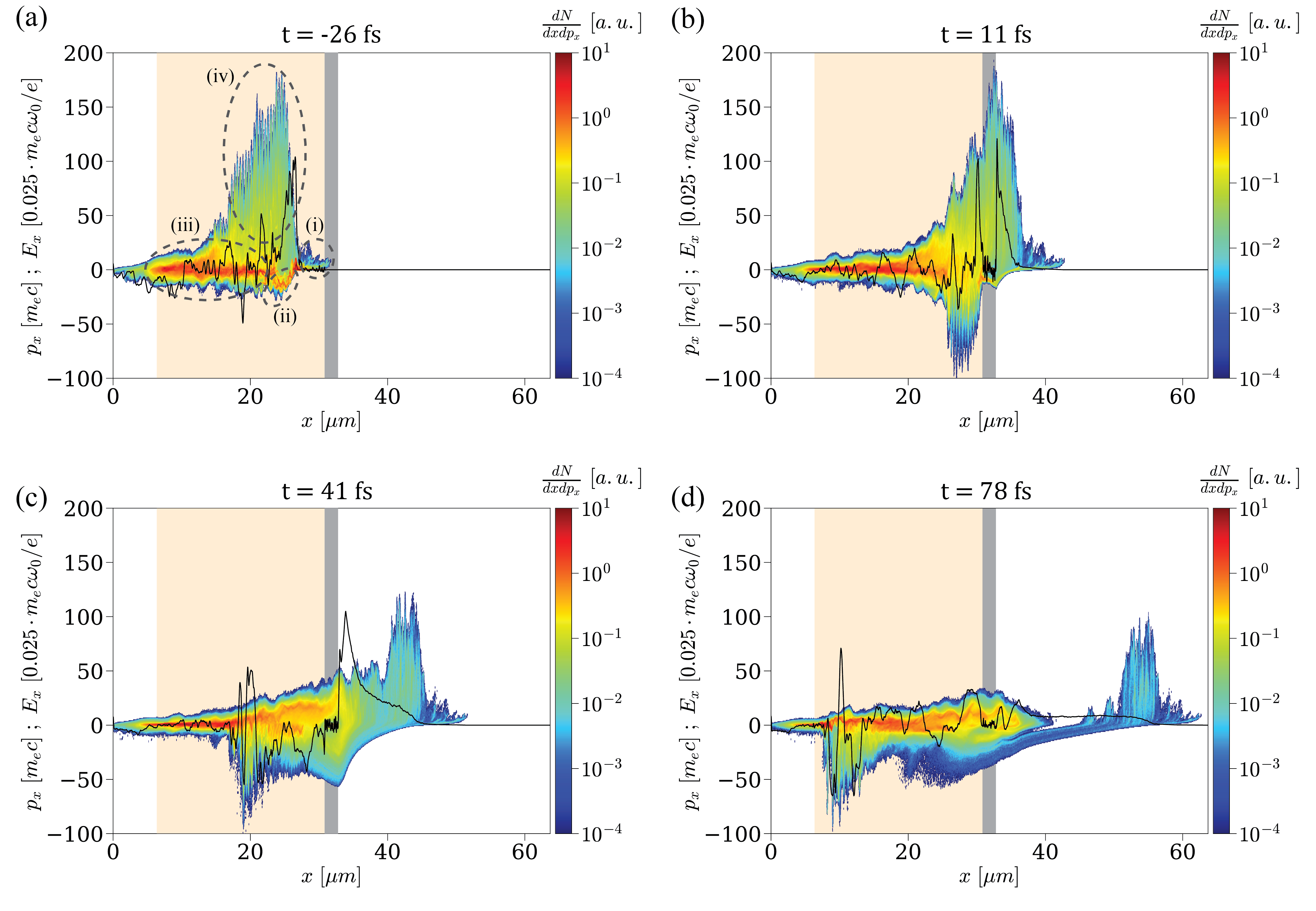} 
\caption{Longitudinal ($x-p_x$) phase space of the gas electrons (colormap) and longitudinal electric field $E_x$ (black curve) along the laser axis ($y=0$) at different times (a)-(d). The gas and foil layers are respectively shown in light orange and gray. Different electron groups are highlighted in (a).}
\label{fig:xpx}
\end{figure*}

Hereafter, the maximum proton energy $\mathcal{E}_{\rm max}$ is defined as the energy obtained in our simulations at $t=400\,\rm fs$. 
One can see in Fig.~\ref{fig:E_max_absorption} that $\mathcal{E}_{\rm max}$ rises with the electron gas density up to a maximum of $\simeq 48\,\rm MeV$ at $n_e=0.5\,n_c$, corresponding to a fourfold increase over the single foil case ($n_e=0$), and decreases at higher densities. To understand this behavior, we plot in Fig.~\ref{fig:E_max_absorption}(b) the variations in $\mathcal{E}_{\rm max}$ and the laser absorption coefficient ($\mathcal{A}$) with the gas density. Interestingly, the proton energy is found to scale about linearly with the laser absorption up to $n_e=0.5n_c$, where the laser energy is almost fully absorbed. 
This linear scaling is not obvious since, assuming a 1D expansion, the maximum proton energy is expected to evolve as $\mathcal{E}_{\rm max} \propto T_h$ up to a logarithm factor depending on $n_h$ and the effective acceleration time \cite{fuchs_laser-driven_2006}, while the laser absorption should scale as $\mathcal{A} \propto n_h T_h \sim n_e T_h$ (taking $n_h \sim n_e$).
As the gas density is further increased up to $2n_c$, the absorption remains nearly saturated while the proton energy is approximately halved.

Since the ion acceleration directly results from the dynamics of the high-energy electrons, we display in Fig.~\ref{fig:xpx}(a-d) the longitudinal ($x-p_x$) phase space of the gas electrons around the laser axis ($y=0$) at different times. At $t=-26\,\rm fs$ [Fig.~\ref{fig:xpx}(a)], the laser pulse has traversed the gas, but has not yet reached the foil. Several distinct electron populations can be identified: (i) a bunch of moderately relativistic ($0 < p_x/m_ec \lesssim 10$) electrons in the foot of the laser pulse ($x\gtrsim 25\,\rm \mu m$); (ii) a dense and relatively cold, backward-moving ($-20\lesssim p_x/m_e c \lesssim 0$) electron flow in the rising part of the laser pulse ($22\lesssim x \lesssim 25\,\rm \mu m$); (iii) a cloud of 
`thermalized', relativistically hot ($\vert p_x \vert/m_e c \lesssim 20$) electrons, traveling in both directions and filling the traversed plasma region; (iv) a spatially modulated cloud of forward-moving, ultra-relativistic ($p_x/m_e c \lesssim 180$) electrons located inside the laser pulse ($15\lesssim x \lesssim 25\,\rm \mu m$).

These electron populations originate from the following sequence of processes.

First, the laser acts as a snowplow when propagating in the gas layer: the electrons accelerated by its ponderomotive force and piled up in its foot make up population (i). As the pulse undergoes erosion during its propagation, its front profile increasingly steepens \cite{vieira_onset_2010}, which enhances the ponderomotive push on the gas electrons
\cite{debayle_electron_2017}.

The charge imbalance due to the electrons piled up ahead of the pulse induces a strong electrostatic field that pulls them back at high negative momenta, resulting in the backward electron flow (ii). At a lower density, this push-pull mechanism would generate an oscillating plasma wakefield. Here, due to the high plasma density, the highly nonlinear wakefield breaks after just one oscillation, thereby converting its kinetic and potential energy into electron heat via phase mixing \cite{debayle_electron_2017}.

Some electrons from this dense thermal reservoir (iii) can be injected into the laser field and driven to ultra-relativistic energies, thus forming population (iv), of same spatial extent as the etched laser pulse. The direct laser acceleration experienced by these electrons is evidenced by their $\lambda_0/2$ ($0.4\,\rm \mu m$) modulations and their maximum momentum ($p_{x,\rm max}/m_e c \simeq 180$), consistent with the prediction $p_{x,\rm max}/m_e c = a_0^2/2$ for a single particle in an electromagnetic plane wave in vacuum. Here account is taken of the intensified laser field (up to $a_0 \simeq 20$) due to self-focusing in the gas. The collective interaction of the fast electrons (iv) with the counterstreaming dense flow (ii) contributes to the fast thermalization of the latter \cite{debayle_electron_2017}. Note that the coupling between the laser and plasma fields may also trigger resonant-type acceleration mechanisms inside the laser pulse \cite{pukhov_particle_1999, zhang_2019}, which can further boost the maximum electron energies.

All of the above energetic electron populations can contribute to TNSA in the foil target. If the laser pulse is able to propagate through the gas layer and reach the foil, the electrons (i) snowplowed at the laser front are injected into the target, and hence induce the rapid early growth of the rear-side TNSA field (see next Section).
In that case, additional MeV-range electrons are generated at the foil's front side through $J\times B$ heating \cite{kruer_jb_1985, debayle_toward_2013}, thus reinforcing the early-time TNSA field. Figure~\ref{fig:xpx}(b) presents the electron phase space
at $t=11\,\rm fs$, when the laser pulse is reflecting off the foil while the ultra-relativistic electrons (iv) are breaking out through its back side, hence amplifying the TNSA field (solid black curve). Some of the bunched electrons (i) have been reflected by the TNSA field, making up most of the backward-moving ($p_x < 0$) electrons visible at the right of the foil.

At $t=41\,\rm fs$ [Fig.~\ref{fig:xpx}(c)], the protons have started expanding from the foil's back side, the fastest ones being located at the peak of the accelerating $E_x$ field ($x=35\,\rm \mu m$). The ultra-relativistic electrons have overtaken the proton front and are being decelerated/reflected by the $E_x$ field (their maximum longitudinal momentum dropping from $p_x/m_e c \simeq 180$ to $\simeq 110$ between $t=11\,\rm fs$ and $t=41\,\rm fs$). A strong $E_x$ field is maintained at the ion front by the hot bulk electrons (iii) that has caught up with the expanding ions. The backward-moving ($p_x/m_e c \gtrsim -70 $) bunch seen at $x\simeq 20\,\rm \mu m$ is made of electrons being accelerated and carried away by the reflected laser pulse.

At $t=78\,\rm fs$ [Fig.~\ref{fig:xpx}(d)], the non-reflected ultra-relativistic electrons have detached themselves from the thermalized bulk electrons that accompany the accelerating ions. These have then attained a maximum energy of $\sim 25\,\rm MeV$ and entered their slower acceleration phase [see Fig.~\ref{fig:E_max_absorption}(a)].

The density dependence of the laser absorption revealed by Fig.~\ref{fig:E_max_absorption}(b) can be explained using the model proposed in Ref.~\cite{debayle_electron_2017}. The typical mean energy ($\simeq T_h$) of the thermal bulk electrons was found to be proportional to the electrostatic potential jump across the nonlinear laser-driven wakefield (modified by the ultrarelativistic electron flow). At moderate laser amplitudes ($a_0 \lesssim 20$), the thermal electron energy weakly depends on the gas density [see Fig.~6 of \cite{debayle_electron_2017}]. Specifically, in the range $12 \le a_0 \le 20$ of interest in our simulations, $T_h$ varies by less than $30\,\%$ when the electron density is increased from $n_e = 0.1$ to $2\,n_c$. Thus, for a fixed gas length, the laser absorption should scale linearly with the gas density, until reaching full absorption when the gas layer is thicker than the laser depletion length. From the model, a 30~fs laser pulse with $12<a_0<20$ is expected to be absorbed after propagating $\sim 25\,\rm \mu m$ in a plasma of density $n_e=n_c$, and $\sim 50\,\rm \mu m$ in a plasma with $n_e = 0.5\,n_c$.

These estimates are consistent with the laser absorption lengths observed in our simulations. For a gas of $0.5\,n_c$ density and $24.5\,\rm \mu m$ length, the laser pulse is still relativistically intense when it hits the foil target, and so the ponderomotively bunched gas electrons and $J\times B$ target electrons can trigger an early proton acceleration. 
At a density $n_e=n_c$, by contrast, the laser pulse is strongly weakened when reaching the foil. Although the electron acceleration by the laser field and plasma wave is similar to that observed at $n_e=0.5n_c$, thus yielding a comparable TNSA field at later times, the absence of the ponderomotively bunched and $J\times B$ electrons suppresses the early rise in the TNSA field, which reduces the final proton energy as shown in Fig.~\ref{fig:E_max_absorption}(b).

\section{Contributions of different plasma regions to the ion-accelerating field}
\label{subsec:contrib_S}

In an electrostatic problem, the electric field at any point in space and time can be calculated as the sum of all electrostatic fields originating from each of the charges in space. Knowing from simulations the spatial distribution of the charge densities, one can calculate the longitudinal electrostatic field arising at $(x,y,t)$ from some region in space $R$ according to the formula
\begin{equation}
E_x(x,y,t)=\frac{1}{2\pi}\int_{R}\frac{\rho(x',y',t)(x-x^\prime)}{(x-x^\prime)^2+(y-y^\prime)^2} dx^\prime dy^\prime
\label{eq:green},
\end{equation}
with $R$ being the plasma region (an area in 2D geometry) of interest.

Using Eq.~\eqref{eq:green}, we investigate the contributions of different plasma regions to the electrostatic field $E_{x,f}(t)$ induced at the on-axis location of the proton front  $(x,y)=(x_{\rm front}(t,0),0)$, where $x_{\rm front}(t,y)$ is the longitudinal position of the outermost (rightmost) protons as a function of time and the transverse coordinate. In practice, $x_{\rm front}(t,y)$ is defined as where the proton density $n_p(x,y)$ vanishes with increasing $x$.

Three plasma regions ($R_i$) are considered: (i) $R_1$ is the region located to the left of the contaminant proton layer; (ii) $R_2$ corresponds to the expanding proton layer; (iii) $R_3$ is the region located to the right of the proton layer.

The boundary between $R_2$ and $R_3$ is readily defined as $x_{2/3}(y) = x_{\rm front}(t,y)$. Obviously, $R_3$ only contains electrons.
Defining the boundary $x_{1/2}(y)$ between $R_1$ and $R_2$ is more complicated as a small fraction of the contaminant protons may initially mingle with the carbon ions and stay inside the carbon plasma at later times. Therefore, looking for the $x$ position where $n_p(x,y)$ starts to rise from 0 is not relevant if one wishes to locate the main accelerated proton population that can break away from the carbon foil [see  Fig.~\ref{fig:density_maps}(b)]. Instead, for each transverse position $y$, we determine the maximum proton density $n_{\rm max}(y) \equiv \mathrm{max}_x n_p(x,y)$ outside of the area where the carbon ions are present, together with its position, $x_{\rm max}(y)$. The boundary between $R_1$ and $R_2$ is taken to be the position $x_{1/2}(y)$ which is the closest (to the left) to $x_{\rm max}(y)$ and satisfies $n_p(x_{1/2})<n_{\rm max}/5$. Note that since the leftmost protons of the contaminant layer can be mixed among the carbon ions (this is particularly true far off axis where the proton layer does not completely detach from the foil's backside), $R_2$ may comprise a few carbon ions and their accompanying electrons.
By contrast, $R_1$ includes almost all carbon ions, the gas protons and all of the electrons located at $x<x_{1/2}$ (including those filling the gap between the carbon ions and the detached proton layer).
The boundaries between the three plasma regions are updated in time, as shown in Fig.~\ref{fig:density_maps} (a,b).

\begin{figure}
\centering
\includegraphics[width=0.8 \columnwidth]{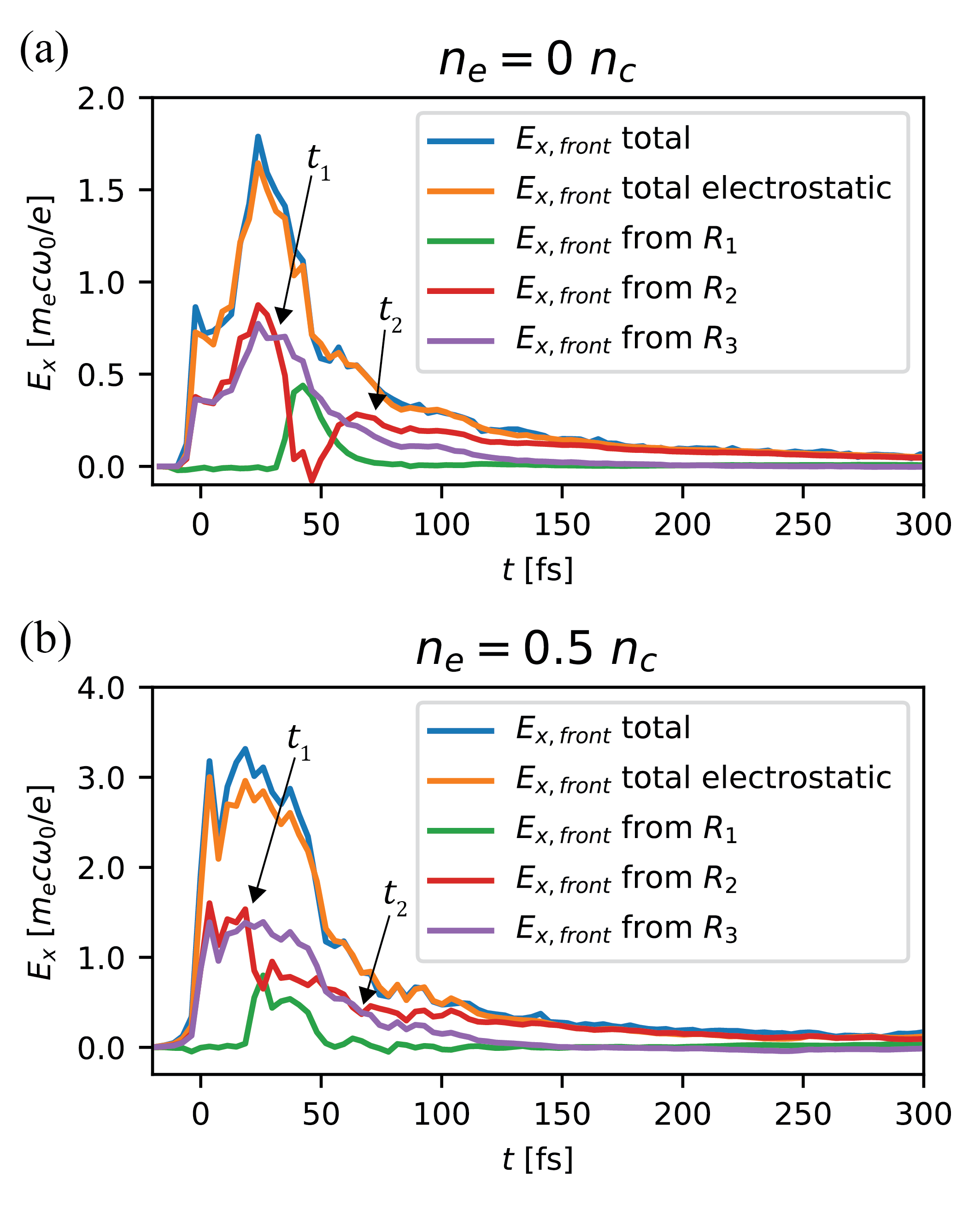}
\caption{Time evolution of the electrostatic fields set up at the proton front by the plasma regions indicated in Fig.~\ref{fig:density_maps}, in the (a) single foil and (b) gas-foil targets. The green, red curve and purple curves plot, respectively, the contributions of the charges lying to the left (region $R_1$), inside (region $R_2$) and to the right (region $R_3$) of the proton layer. These fields have been computed using Eq.~\eqref{eq:green}. The total electrostatic field (orange) is also plotted and compared with the total electric field measured in the simulation. The times $t_1$ and $t_2$ correspond to the detachment of the proton layer from the foil and the onset of lateral effects, respectively.}
\label{Ex_LCR}
\end{figure}

The time evolution of the electrostatic fields generated at the proton front $x_{\rm front}(t,0)$ by the three plasma regions is presented in Fig.~\ref{Ex_LCR}(a) for the foil target and in Fig.~\ref{Ex_LCR}(b) for the gas-foil target with $n_e=0.5\,n_c$. Moreover, in order to confirm the validity of the electrostatic approximation, the total electrostatic field (orange curve), obtained by summing up the contributions of the three regions, is compared to the total electric (not just electrostatic) field extracted from the simulation (blue curve). The observed good agreement between the two fields validates the electrostatic approximation, and hence the use of Eq.~\eqref{eq:green} to estimate the TNSA field.

Although reaching a $\sim 80\,\%$ stronger maximum in the gas-foil target, $E_{x,\rm front}(t)$ behaves similarly in both targets. Early in time, the dominant contributions to $E_{x,\rm front}$ come from the electrons in $R_3$ (purple curve) and the net positive charge contained in $R_2$ (red curve). These two regions generate positive (ion-accelerating) electric fields of same amplitude which add up. In other terms, the rightmost (fastest) protons are as much pulled by the sheath electrons in $R_3$ as they are pushed by the positively charged proton layer covered by $R_2$. This is an expected result in the 1D geometry that characterizes the initial ion expansion. By contrast, the carbon ions and electrons from $R_1$ produce a negligible net field at the proton front. In both targets, the field shows a sudden rise to reach a first peak around $t\simeq 0$. As discussed above, it is caused by the arrival of the snowplowed electrons from the gas and the $J\times B$ heated electrons from the foil surface.

At $t=t_1$ [see Fig.~\ref{Ex_LCR}(a,b)], the heated carbon plasma starts to expand, leading to a dense electron sheath permeating the leftmost, lagging part of the proton layer. This transiently reduces the field from $R_2$, while, in return, the positive net charge of $R_1$ produces a significant accelerating field. This effect, which appears to be most pronounced in the single foil target, ceases when most of the protons have detached from the foil.

At $t=t_2$ [see Fig.~\ref{Ex_LCR}(a,b)], the on-axis part of the proton layer is well separated from the carbon plasma, so that the fields from $R_2$ and $R_3$ again mainly account for the total accelerating field. Yet from this time onwards,
the push from the positively charged proton layer ($R_2$) begins to dominate the fast-decreasing pull from the sheath electrons ($R_3$). Actually, the field from the sheath electrons diminishes up to the point of even reaching slightly negative values by $t\simeq 150\,\rm fs$. This behavior results from a multidimensional effect, namely, the transverse curvature of the proton front as evidenced in Fig.~\ref{fig:density_maps}(b). The rightmost protons on axis are overtaking most of the off-axis sheath electrons, which end up exerting a decelerating force on those protons. The acceleration of the fastest protons is then solely driven by the repulsive positive charge of the proton layer.

\section{Hot-electron contributions to the ion-accelerating field}

A closer look at the temporal evolution of the total accelerating field [Fig.~\ref{Ex_LCR}] allows one to understand why the gas-foil target performs better than the single foil target. Firstly, the field reaches its absolute maximum much faster and this maximum is also about $1.8\times$ higher. Secondly, the field strength remains close to its peak value during $\sim 50\,\rm fs$, which differs from the rapid post-peak decay observed with the single foil. Since the final proton energy is proportional to the square of the integral of the field curve, the above observations explain the much improved proton acceleration achieved in the gas-foil target (see Fig.~\ref{fig:E_max_absorption}).

In the single foil target, the rise time of the accelerating field is consistent with the 30-fs duration of the laser pulse. Its subsequent drop is caused by the plasma expansion and the transverse dilution of the hot electrons.
The higher field maximum and subsequent plateau seen in the gas-foil target are due to the successive exits across the foil back side of the distinct hot-electron groups issued from the laser-gas and laser-foil interactions.

\begin{figure}
\centering
\includegraphics[width=0.8\columnwidth]{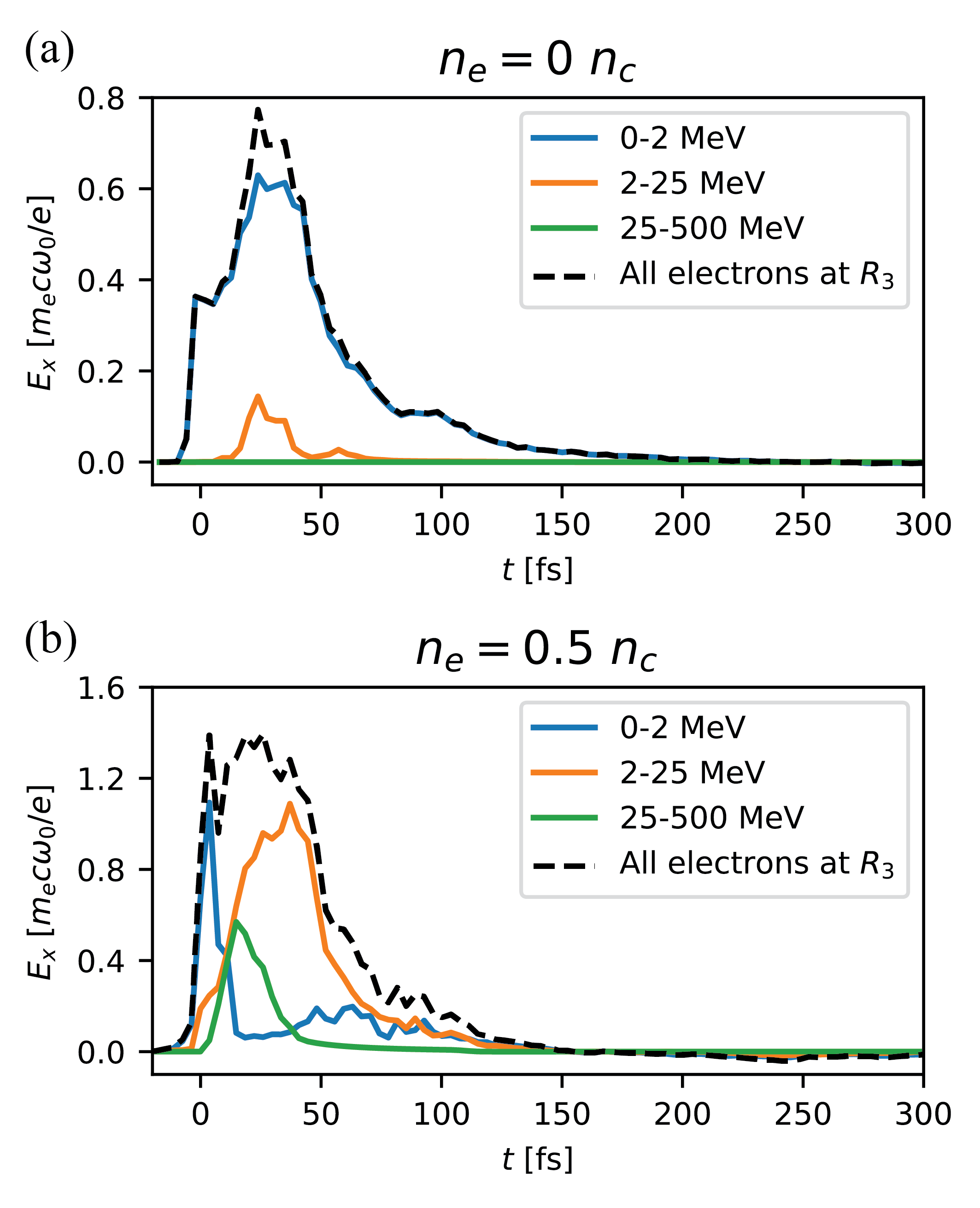}
\caption
{
\label{Ex_populations}
Time evolution of the electric fields generated by different electron groups (see legend) located to the right of the proton front in the (a) single and (b) gas-foil targets. The dashed black curve plots the total electric field due to all of the electrons lying in region $R_3$.
}
\end{figure}

To elucidate these different electron contributions, we divide the electron population lying in region $R_3$ into three kinetic energy ranges: From zero to about the ponderomotive energy ($<2\,\rm MeV$), medium energy ($2-25\,\rm MeV$) and high energy ($>25\,\rm MeV$). We \textcolor{black}{consider} only the electrons to the right of the front ($x> x_{\rm front}(t,0)$) since they contribute positively to the proton acceleration at all times (unlike the far-axis electrons as discussed above). Again using Eq.~\ref{eq:green}, we plot in Figs.~\ref{Ex_populations}(a,b) the electrostatic field generated at the proton front by each electron group in the single foil (a) and gas-foil (b) targets.

With a single foil target [Fig.~\ref{Ex_populations}(a)], the accelerating field is mainly generated by low-energy electrons ($<2\,\rm MeV$), produced by $J\times B$ heating~\cite{kruer_jb_1985, debayle_toward_2013}. The electrons in the $2-25\,\rm MeV$ energy range contribute to only $\sim 10\,\%$ of the field, while the ultra-relativistic ($>25\,\rm MeV$) electrons play a negligible role. 

In the gas-foil target [Fig.~\ref{Ex_populations}(b)], the sheath field results from more diverse electron sources. The $<2\,\rm MeV$ electrons snowplowed by the laser in the gas or $J\times B$ accelerated at the foil surface are responsible, approximately equally, for the early peak of the sheath field, which lasts for $\sim 15\,\rm fs$. Higher-energy electrons from the gas then take over. 
The fastest ($25-500\,\rm MeV$) electrons momentarily accounts for $\sim 40\,\%$ of the total field, but their contribution rapidly weakens as they move away from the proton front (to a distance $r$ larger than their transverse extent, causing the field to decrease as $\sim 1/r$ in 2D). Actually, from $t \simeq 15\,\rm fs$, most of the field originates from the thermal mid-energy ($2-25\,\rm MeV$) electrons. As they extend longitudinally throughout the gas-foil target [see Fig.~\ref{fig:xpx}(d)], they make up a large reservoir of kinetic energy, capable of sustaining a strong sheath field for a (relatively) long time. As they lose energy to the expanding ions, the lower-energy electrons from this population progressively fall into the $<2\,\rm MeV$ group, which leads to the associated field to rise again by $t\simeq 50\,\rm fs$. Later in time ($t\gtrsim 70\,\rm fs$), the $<2\,\rm MeV$ and $2-25\,\rm MeV$ electron groups turn out to contribute similarly to the sheath field.

\section{Conclusions}

Using PIC simulations, we have investigated in detail the processes leading to efficient TNSA in gas-foil targets irradiated by a $3\times 10^{20}\,\rm Wcm^{-2}$, 30~fs laser pulse. 
For a fixed gas length of $\sim 25\,\rm \mu m$, we have found an optimum hydrogen gas density of $0.5\,n_c$. These parameters entail an almost complete laser absorption through a combination of electron energization mechanisms involving the laser field and the excited nonlinear plasma waves. Their interplay generates several energetic electron populations which differ by their phase-space properties, each of these populations accounting significantly for proton acceleration in the solid foil.

The main electron contribution to the TNSA field comes from the relativistically hot ($\sim 10\,\rm MeV$) electron bulk resulting from the laser-gas interaction, which generate a strong and relatively long-lived sheath field. The ultra-relativistic gas electrons directly accelerated by the laser field produce a substantial field too, but this field decays quite rapidly as they move away from the foil.
The lower-energy ($\sim \rm MeV$) electrons snowplowed by the laser in the gas or driven at the foil surface play also a crucial role in building the sheath field at early time and pre-accelerate the protons, resulting in an improved overall energy gain. These MeV electrons are efficiently generated only if the laser is intense enough when reaching the foil. The optimal configuration thus corresponds to the case where the laser energy is nearly fully converted into electron kinetic energy inside the gas, but is still able to drive electrons to relativistic energies at the foil surface.

Moreover, this study allowed us, for the first time, to identify as a function of time the plasma regions responsible for the sheath field seen by the fastest protons. This was done by reconstructing, in the electrostatic limit, the electric field at the proton front from the surrounding charge density distributions. At early times, proton acceleration takes place in an essentially 1D geometry, so that the accelerating field is produced equally by the electrons to the right of the proton front and the net positive charge contained in the proton layer. By contrast, after $\sim 70\,\rm fs$, the proton front develops a curvature due to the transverse proton velocity gradient, which causes the electrons to exert a negligible net force on the outermost protons. The subsequent acceleration of the latter is then mostly sustained by the continuous push exerted by the positively charged proton layer. This new method of analysis of the ion-accelerating field can be readily generalized to other target setups.

In our simulations, only the gas density was modified. In general, however, proton acceleration will depend on additional parameters such as gas length, foil thickness, laser intensity and duration and on the location of the focal plane. To aid the determination of the optimal gas parameters, which correspond to almost full laser absorption through the gas, one can make use of the scaling given in Ref.~\cite{debayle_electron_2017}.

To conclude, we have demonstrated that gas-foil targets constitute a promising configuration for TNSA-type proton acceleration under laser interaction conditions available in many university-scale laboratories.

\section*{Acknowledgments}
D. L. and V. M. were
supported by research grants from the Wolfson Foundation and from Dita and Yehuda L. Bronicki. 

\bibliography{bibliography}
\bibliographystyle{apsrev4-1}

\end{document}